%% file: main.tex
\documentclass[default]{sn-jnl}% Default
\usepackage{amsmath,amsfonts,amssymb}
\usepackage{graphicx}
\usepackage{adjustbox}
\usepackage{graphicx}
\usepackage{siunitx}
\usepackage{mathtools}
\usepackage{threeparttable}
\usepackage{float}
\usepackage{booktabs}
\usepackage{tabularx}
\usepackage{makecell}
\newcolumntype{C}[1]{>{\centering\arraybackslash}p{#1}}
\newcolumntype{L}[1]{>{\raggedright\arraybackslash}p{#1}}
\usepackage{multicol}
\usepackage[T1]{fontenc}
\usepackage{csquotes}
\usepackage{url}
\usepackage{natbib}
\usepackage{caption}
\usepackage{subcaption}
\usepackage{xcolor}
\usepackage{xr}
\usepackage{cleveref} %% for the supplementary material tex file -- use package to link content in suppMat.tex in main.tex

\graphicspath{ 
{fig_money_model_arch/} 
{fig_results/}
{fig_exampleCT/}
}

\begin{document}

\title[Anatomical Prior Guided Segmentation of Mediastinal Lymph Nodes in CT]{Segmentation of Mediastinal Lymph Nodes in CT with Anatomical Priors}

\author*[1]{\fnm{Tejas Sudharshan} \sur{Mathai}}
\author[2]{\fnm{Bohan} \sur{Liu}}
\author[1]{\fnm{Ronald M.} \sur{Summers}}

\affil[1]{\orgdiv{Clinical Center, National Institutes of Health (NIH)}, \orgaddress{\state{MD}, \country{USA}}}
% \affil[2]{\orgdiv{National Library of Medicine}, \orgname{NIH}, \orgaddress{\state{MD}, \country{USA}}}
\affil[2]{Department of Radiology, School of Medicine and Health Sciences, George Washington University, \orgaddress{\state{Washington DC}, \country{USA}}}

% \affil[1]{\orgdiv{Imaging Biomarkers and Computer-Aided Diagnosis Laboratory, Clinical Center}, \orgname{NIH}, \orgaddress{\city{Bethesda}, \state{MD}, \country{USA}}}

% \affil[2]{\orgdiv{National Center for Biotechnology Information, National Library of Medicine}, \orgname{NIH}, \orgaddress{\city{Bethesda}, \state{MD}, \country{USA}}}

% ================================================

% ================================================
% Abstract
% ================================================

\abstract{
\textbf{Purpose:} Lymph nodes (LNs) in the chest have a tendency to enlarge due to various pathologies, such as lung cancer or pneumonia. Clinicians routinely measure nodal size to monitor disease progression, confirm metastatic cancer, and assess treatment response. However, variations in their shapes and appearances make it cumbersome to identify LNs, which reside outside of most organs.  

\textbf{Methods:} We propose to segment LNs in the mediastinum by leveraging the anatomical priors of 28 different structures (e.g., lung, trachea etc.) generated by the public TotalSegmentator tool. The CT volumes from 89 patients available in the public NIH CT Lymph Node dataset were used to train three 3D nnUNet models to segment LNs. The public St. Olavs dataset containing 15 patients (out-of-training-distribution) was used to evaluate the segmentation performance. 

\textbf{Results:} For the 15 test patients, the 3D cascade nnUNet model obtained the highest Dice score of 72.2 $\pm$ 22.3 for mediastinal LNs with short axis diameter $\geq$ 8mm and 54.8 $\pm$ 23.8 for all LNs respectively. These results represent an improvement of 10 points over a current approach that was evaluated on the same test dataset. 

\textbf{Conclusion:} To our knowledge, we are the first to harness 28 distinct anatomical priors to segment mediastinal LNs, and our work can be extended to other nodal zones in the body. The proposed method has immense potential for improved patient outcomes through the identification of enlarged nodes in initial staging CT scans. 

% Our findings indicated that anatomical priors were helpful for nnUNet to segment mediastinal nodes, and can be extended to other nodal zones in the body.
}

\keywords{CT, Lymph Node, Mediastinum, Segmentation, Anatomical Priors, Deep Learning}

\maketitle

% ================================================

% ================================================
% Chapters
% ================================================

\input{chapters/1_introduction.tex}
\input{chapters/2_methods.tex}
\input{chapters/3_experiments_and_results.tex}

\input{chapters/4_discussion_and_conclusion.tex}
\input{chapters/5_acknowledgements.tex}

% ================================================

% \bibliography{sn-bibliography}% common bib file
% %% if required, the content of .bbl file can be included here once bbl is generated
% % \input sn-article.bbl

% %% Default %%
% % \input sn-sample-bib.tex%

% {
% \footnotesize
% \bibliography{sn-bibliography}
% }

\clearpage
{\tiny \bibliography{main.bib}}
% {\tiny \bibliography{main}}
\bibliographystyle{unsrt}
\clearpage
% ================================================

%% include supplementary material
\input{chapters/supplementaryMaterial.tex}

\end{document}

%% file: chapters/1_introduction.tex
% ================================================
\section{Introduction}
\label{sec_Introduction}
% ================================================

Lymph nodes and the lymphatic system comprise an integral part of the body's natural defense mechanisms and play a vital role in maintaining a person's health. Abnormalities to the lymphatic system can result in enlarged lymph nodes (lymphadenopathy) \cite{Torabi2004,Ganeshalingam2009} with etiologies ranging from infection, autoimmune disease or malignancy. Distinguishing between the causes for enlarged and metastatic nodes from non-metastatic LNs is critical for clinicians in determining the correct treatment \cite{Torabi2004,Ganeshalingam2009,Taupitz2007,Amin2017}. Frequently, radiologists use a systematic approach to identify suspicious nodes through nodal size measurement with the help of established guidelines, such as the tumor, node, and metastasis (TNM) criteria \cite{Amin2017}. In particular, the presence of enlarged LNs in the setting of cancer not only dictates the staging and extent, but is vital to treatment and management. 

In clinical practice, radiologists routinely identify, manually measure, and describe the features of lymph nodes on CT and MRI to identify areas of pathology. Among the various imaging features for lymphadenopathy, nodal size is the most widely used criteria \cite{Torabi2004,Ganeshalingam2009,Taupitz2007,Amin2017} to determine benign versus malignant status when paired with clinical data. A node is considered enlarged if its smallest diameter (along the short axis) is greater than 10mm on an axial CT slice \cite{Torabi2004,Ganeshalingam2009,Taupitz2007,Amin2017,Mao2014}. However, this assessment can be cumbersome and time-consuming, especially at initial staging and while comparing multiple sites of metastasis during the evaluation of treatment response in follow-up imaging. To help relieve this laborious process, automated LN measurement can augment radiology workflows by aiding in the identification of LNs in specific regions of the body, such as the mediastinum. 

Several approaches have been proposed to detect and segment mediastinal lymph nodes in both CT \cite{Liu2014_spatialPrior,Roth2014_NIHCTLN,Oda2018,Bouget2019_StOlavs_15Patients,Iuga2021_fovealNetwork_NIHCTLN,Iuga2021_stationMapping_NIHCTLN,Bouget2022_StOlavs,Mehrtash2023} and MRI \cite{Lu2018_mri,Debats2019_mri,Mathai2021_mlmi,Mathai2021_spie,Wang2022_mri,Mathai2022_CARS,Mathai2023_LNMRISeg}. Only a handful \cite{Liu2014_spatialPrior,Bouget2019_StOlavs_15Patients,Bouget2022_StOlavs,Iuga2021_stationMapping_NIHCTLN,Oda2018,Mehrtash2023} exploit the anatomical prior information that plays a significant role in reducing the number of false positives through disambiguation of collocated lymph nodes and other structures of similar intensity (e.g., esophagus, azygos vein). Presently, only 4 anatomical regions have been used in prior works \cite{Bouget2022_StOlavs,Oda2018} to distinguish mediastinal LNs from other adjacent structures. We are the first to segment mediastinal nodes by leveraging the anatomical priors of 28 different structures in the body, and thereby account for the aforementioned challenges in the radiology workflow.

% ^^^^^^^^
% ^^^^^^^^
\begin{figure*}[!tb]
\centering
\includegraphics[width=\textwidth]{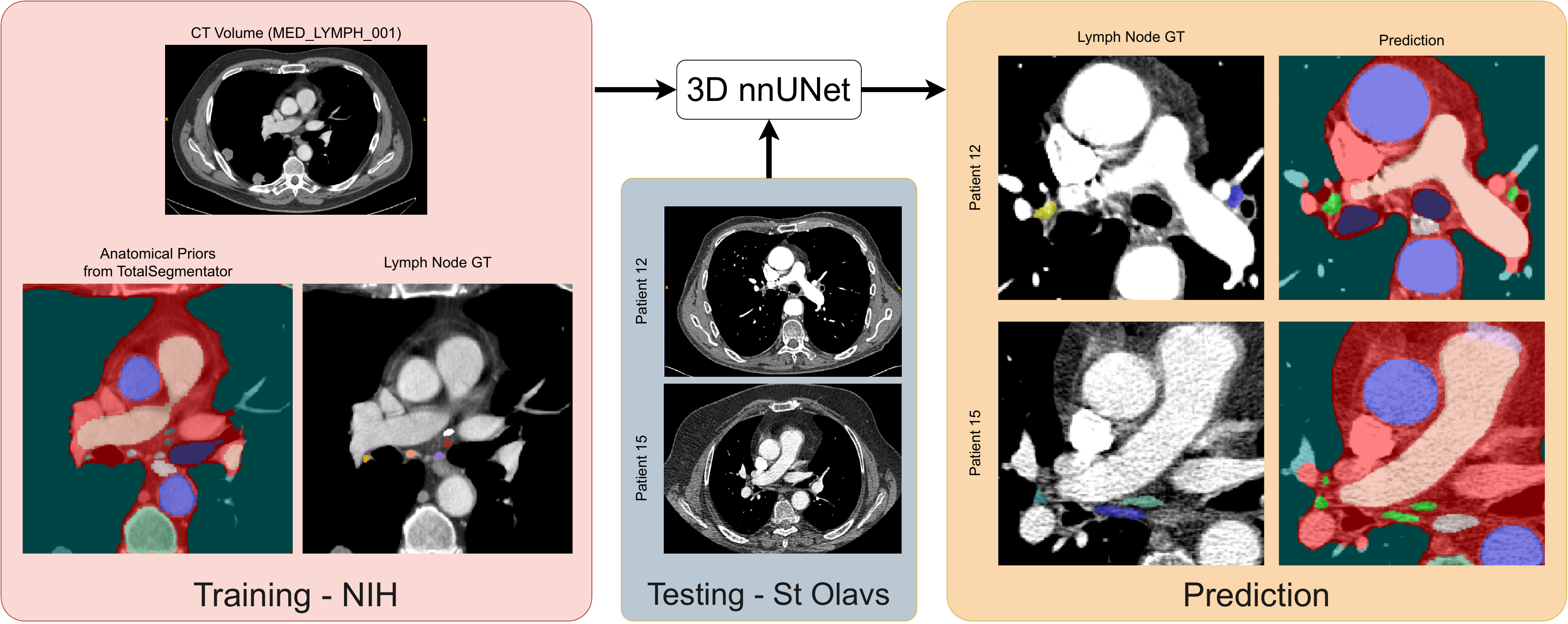}
\caption{Flowchart of the proposed approach to segment mediastinal lymph nodes in CT using anatomical priors. First, the public TotalSegmentator tool was used to segment 28 structures in 89 mediastinal CT volumes from the public NIH CT Lymph Node dataset. Next, these labels were combined with the manual annotations for mediastinal LNs, and used to train a 3D nnUNet segmentation model. At test time, the 3D nnUNet was executed on CT volumes of 15 patients in the public St Olavs dataset. Green labels in the prediction correspond to the predicted LNs.     The figure is best viewed in color in the PDF.}
\label{fig_money}
\end{figure*}
% ^^^^^^^^
% ^^^^^^^^

In this paper, we present an approach to segment mediastinal LNs in CT studies of the body. Fig. \ref{fig_money} shows an overview of the pipeline. We used the LN labels for 89 CT volumes from the public NIH CT Lymph Node dataset, and combined them with the labels for 28 distinct structures in the body obtained through the public TotalSegmentator tool. Three nnUNet segmentation models were trained end-to-end with this data, and evaluated on a test dataset comprising of 15 patients from an external institution. Our results indicated a performance improvement (measured through Dice scores) over the current state-of-the-art method evaluated on the same test dataset. 

%% file: chapters/2_methods.tex
% ================================================
\section{Methods}
\label{sec_Methods}
% ================================================

\subsection{Patient Population.} 

We used datasets from two distinct institutions for the purposes of training and testing the 3D nnUNet models. The public NIH CT Lymph Node dataset \cite{Roth2014_NIHCTLN,Clark2013_TCIA} was used for training, and it comprised of a total of 176 CT series from 176 patients. 90 CT volumes were obtained at the level of the chest (mediastinum), and segmentation masks for 388 nodes with a short axis diameter (SAD) $\geq$ 1cm were provided. The remaining 86 CT volumes were acquired at the abdomen with 595 abdominal LNs annotated. To our knowledge, no underlying disease causes or demographics were provided for the patients in the NIH dataset, and LNs that were smaller than 1cm were left unannotated. 

However upon visual inspection, only 89 of the 90 mediastinal CT volumes had a field-of-view centered around the thorax \cite{Bouget2022_StOlavs}. Additionally, Bouget et al. \cite{Bouget2022_StOlavs} provided the ground truth annotations for all the mediastinal LNs in these 89 volumes. In particular, the authors adopted a ``conservative'' annotation approach and segmented all suspicious regions as lymph nodes. They incorporated nodes with any short-axis measurement including nodes smaller than the suggested RECIST criterion for malignancy of 1cm. We used these labels for the 89 volumes for training our models. 

The test dataset was a public dataset released from St Olavs Hospital in Trondheim, Norway \cite{Bouget2019_StOlavs_15Patients,Reynisson2015_StOlavs_lungCancerPatientsOriginalData} that comprised of 15 patients with confirmed lung cancer diagnosis. A total of 384 lymph nodes were annotated in this dataset with 131 nodes having a SAD $\geq$ 8mm and 238 nodes with SAD $<$ 8mm. The dimensions of the CT volumes in the test set ranged from (487 $\sim$ 512) $\times$ (441 $\sim$ 512) $\times$ (241 $\sim$ 829) voxels. To our knowledge, this is the largest publicly available dataset in which all mediastinal LNs have been segmented. Since LNs with a SAD $\geq$ 8mm are suspicious for metastasis \cite{Taupitz2007,Mathai2023_LNMRISeg,Mathai2022_CARS}, we considered these nodes as clinically significant in this work. Furthermore, in contrast to prior works, we focused solely on LN segmentation and not on station mapping.

\subsection{Anatomical Priors} 

Inspired by prior literature \cite{Liu2014_spatialPrior,Bouget2019_StOlavs_15Patients,Bouget2022_StOlavs,Iuga2021_stationMapping_NIHCTLN,Oda2018,Mehrtash2023} on LN segmentation with anatomical priors, we utilized the public TotalSegmentator \cite{Wasserthal2023_TS} that was designed to segment over 117 distinct classes in CT volumes. The tool is of tremendous use for various applications, such as personalized risk assessment through body composition analysis \cite{Hou2024_bodyComposition,Lee2023_bodyCompositionAnalysis}. TotalSegmentator was developed using a training set of 1,204 CT exams and encompasses a diverse array of scanners, institutions, and protocols to ensure its versatility and robustness in different clinical settings. We utilized the segmentation labels generated by this tool for 28 different structures (e.g., trachea, pulmonary artery etc.) in the body and combined them with the lymph node labels, resulting in a total of 29 distinct classes for training. A complete list of the segmentation classes provided by TotalSegmentator is presented in Supplementary Material Table \ref{fig:model_selection_criteria}. Incorporation of anatomical priors helped to disambiguate anatomical regions of the body that are of similar intensity as the LNs, such as the heart and esophagus, and it also reduced the effect of severe class imbalances between the lymph node labels (foreground) and the background. Furthermore, we were solely interested in the segmentation of LNs, so we only used the 28 classes for training and discarded them at test time.

\subsection{3D nnUNet} 

The self-configuring nnUNet segmentation framework \cite{Isensee_2021} was employed to train different configurations for the task of LN segmentation in CT. The nnUNet model is currently the \textit{de-facto} standard for segmentation, and it can be adapted for various datasets and modalities, including CT and MRI. The framework automatically determined the optimal hyper-parameters for training a segmentation model and learned to segment target structures of interest. In this work, we trained 3D low-resolution, 3D full-resolution, and 3D cascade nnUNet configurations, and compared their performance. 

During training, each configuration of the 3D nnUNet took as input the CT (unwindowed) volume and the corresponding ground-truth masks for 29 different structures. Five-fold cross-validation with different initialization of trainable parameters for a total of 1000 epochs was done. Distinct subsets of training and validation data from the 89 CT volumes were automatically created for each fold. The model learned to segment the target structures of interest in the volume, and iteratively refined it via a loss function. The loss function used by the model was an equally weighted combination of binary cross-entropy and soft Dice losses. This loss function computed a segmentation error that measured the overlap between the prediction and ground-truth. It was optimized using the Stochastic Gradient Descent (SGD) optimizer with an initial learning rate of $10^{-2}$ and a batch size of 1.

At test time, the 3D nnUNet predicted the segmentation masks for the structures in the held-out test CT volumes. Since our primary objective was to segment LNs, we discarded the remaining classes at test time. The best model with the lowest loss from each of the 5 folds was used for inference on the test CT volume, and predictions from these five folds were ensembled together. All experiments were run on a desktop running Ubuntu 20.4 LTS with a NVIDIA V100 GPU with 32GB RAM.

%%%%%%%%%%%%%%%%%%%%%%%%%%%%%%%%%%%%%%%%%%%%%%%%%%%%%%%%%%%%%%%%%%
\begin{figure}[!htb]
    \centering
    \begin{subfigure}[b]{\textwidth}
        \centering
        \includegraphics[width=0.3\textwidth]{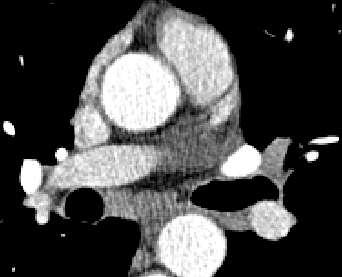}%
        \hfill
        \includegraphics[width=0.3\textwidth]{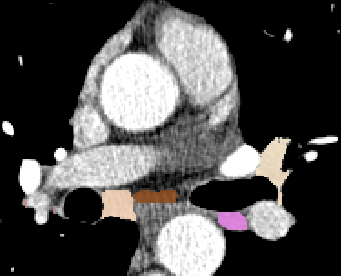}
        \hfill
        \includegraphics[width=0.3\textwidth]{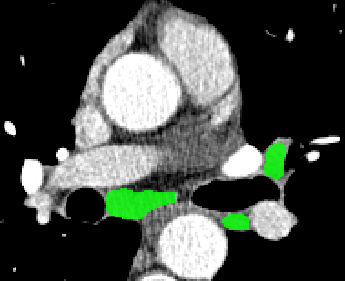}
        \caption{Patient 1}
    \end{subfigure}
    
    \begin{subfigure}[b]{\textwidth}
        \centering
        \includegraphics[width=0.3\textwidth]{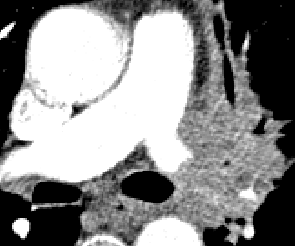}%
        \hfill
        \includegraphics[width=0.3\textwidth]{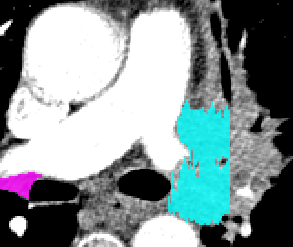}
        \hfill
        \includegraphics[width=0.3\textwidth]{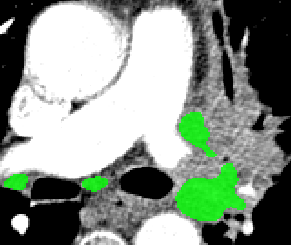}
        \caption{Patient 7}
    \end{subfigure}
    
    \begin{subfigure}[b]{\textwidth}
        \centering
        \includegraphics[width=0.3\textwidth]{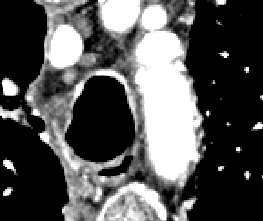}%
        \hfill
        \includegraphics[width=0.3\textwidth]{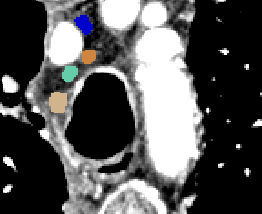}
        \hfill
        \includegraphics[width=0.3\textwidth]{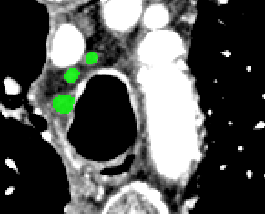}
        \caption{Patient 10}
    \end{subfigure}
    \caption{Results from our approach to detect mediastinal LNs in CT volumes. Left column: A slice of the original CT volume, Middle column: GT annotation, Right column: Prediction from the nnUNet Cascade model. The different colors in the GT correspond to the different stations of the LNs, but for evaluation purposes, they were all considered to belong to one class based on their short axis diameter. Notice that in (b) for patient \#7, the model was able to partially capture the large metastatic node (blue), while it also identified an unmarked node in the GT (middle). In (c), the model missed the node in blue.}
    \label{fig_example_CT_predictions}
\end{figure}
%%%%%%%%%%%%%%%%%%%%%%%%%%%%%%%%%%%%%%%%%%%%%%%%%%%%%%%%%%%%%%%%%%

%% file: chapters/3_experiments_and_results.tex
% ================================================
\section{Results}
\label{sec_exp} 
% ================================================

\subsection{Experiments and Metrics} 

The 3D nnUNet models in our work were trained with the data acquired at the NIH and tested on data obtained at an external institution (out-of-training distribution). Therefore, our primary experiment was the comparison against the slab-wise UNet designed by Bouget et al. \cite{Bouget2022_StOlavs}, which was evaluated on the same test dataset of 15 patients. Additionally, we also wanted to determine the performance of the individual configurations of the 3D nnUNet on the test set. In order to quantify this, we used the Dice coefficient as the metric. Contrary to prior works \cite{Bouget2019_StOlavs_15Patients,Bouget2022_StOlavs,Mehrtash2023}, we did not apply any post-processing techniques to the predictions from our segmentation models. We only partitioned the LNs based on their SAD, and computed results for each component in the division. 

% ^^^^^^^^
% ^^^^^^^^
\begin{table*}[!ht]
\centering\fontsize{9}{12}\selectfont % to make font size 9 pt
\setlength\aboverulesep{0pt}\setlength\belowrulesep{0pt} % intersect vert and horiz lines
\setlength{\tabcolsep}{7pt} % set small spacing between entries of column (default 6pt)
\setcellgapes{3pt}\makegapedcells % small space between row entries (default 3pt)
\caption{Comparison of the different nnUNet models for the LN segmentation task. Bold font indicates best results. ``-'' stands for unreported results. }
\begin{adjustbox}{max width=\textwidth}
\begin{tabular}{@{} c|c|c|c|c @{}} % @ property can be modified
\toprule
\#      &   Method      & Dice (LN $\geq$ 8mm)      & Dice (LN $<$ 8mm)         & Dice (All LN) \\
\midrule

1       &   Bouget et al. 2022      & -         & -         & 44.8 $\pm$ 13.5  \\

\midrule

2       &   3D nnUNet (Low Res)                    & 57.7 $\pm$ 28.0         & 29.1 $\pm$ 21.2         & 47.3 $\pm$ 24.3  \\
3       &   3D nnUNet (Full Res)                   & 65.8 $\pm$ 24.8         & 41.4 $\pm$ 22.6         & 52.0 $\pm$ 25.4  \\
4       &   3D nnUNet (Cascade \& Low Res)         & 68.6 $\pm$ 20.6         & 38.9 $\pm$ 23.4         & 54.5 $\pm$ 23.2  \\
4       &   3D nnUNet (Cascade \& Full Res)        & \textbf{72.2 $\pm$ 22.3}         & \textbf{49.6 $\pm$ 23.5}         & \textbf{54.8 $\pm$ 23.8}  \\

\bottomrule
\end{tabular}
\end{adjustbox}
\label{table_LN_dice_comparisons}
\end{table*}
% ^^^^^^^^
% ^^^^^^^^

%%%%%%%%%%%%%%%%%%%%%%%%%%%%%%%%%%%%%%%%%%%%%%%%%%%%%%%%%%%%%%%%%%
\begin{figure}[!ht]
\centering
\includegraphics[width=0.85\columnwidth]{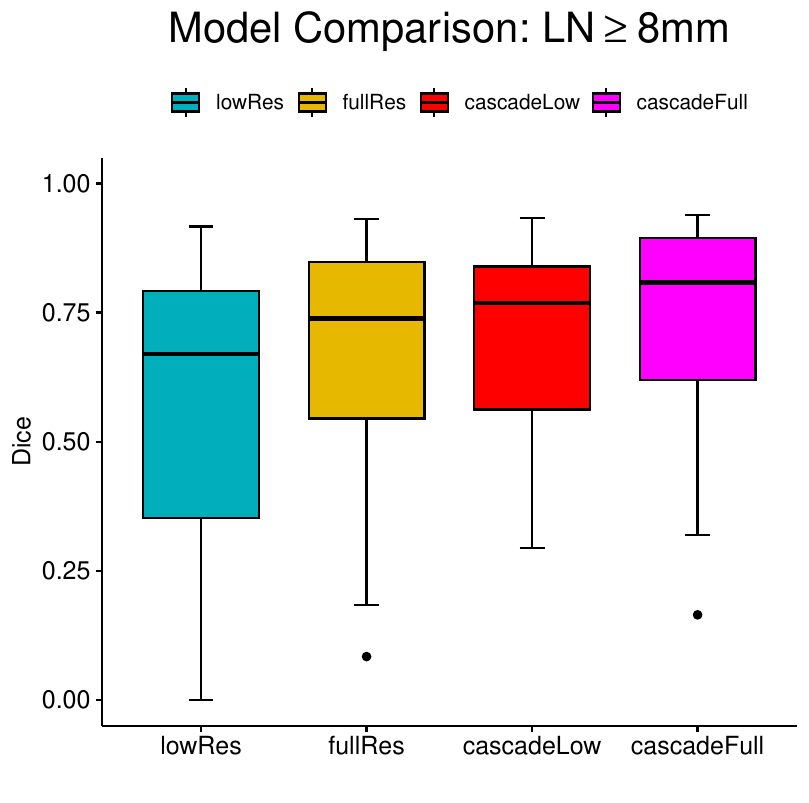}
\caption{Box plots of the different 3D nnUNet model configurations for the segmentation of mediastinal lymph nodes in the St Olavs dataset. Results are shown for lymph nodes with short axis diameters $\geq$ 8mm.}
\label{fig_box_plots}
\end{figure}
%%%%%%%%%%%%%%%%%%%%%%%%%%%%%%%%%%%%%%%%%%%%%%%%%%%%%%%%%%%%%%%%%%

\subsection{Results} 

Table \ref{table_LN_dice_comparisons} provides a summary of the Dice scores across the different 3D nnUNet configurations. It also shows our model's performance in contrast to prior work \cite{Bouget2022_StOlavs} on the same test set. Fig. \ref{fig_box_plots} contains box plots that shows the distribution of the Dice scores across the different nnUNet configurations. It can be seen from the table that the 3D low-resolution nnUNet fares the worst amongst all the configurations. The 3D full-resolution nnUNet yielded respectable Dice scores across all LNs and for those LNs with SAD $\geq$ 8mm. However, the 3D cascade nnUNet models performed the best. Of note, the 3D cascade nnUNet with the first-stage predictions from the full-resolution model demonstrated the best Dice scores across all LN size categories. This model achieved a Dice score (mean and standard deviation) of 54.8 $\pm$ 23.8 across all LNs regardless of their short axis diameters, 49.6 $\pm$ 23.5 for LNs with SAD $<$ 8mm, and 72.2 $\pm$ 22.3 for LNs with SAD $\geq$ 8mm, respectively. In contrast to the Dice score of 44.8 $\pm$ 13.5 obtained by Bouget et al. \cite{Bouget2022_StOlavs}, our results show a marked improvement with a 10 point increase in Dice. These results are corroborated by the box plots in Fig. \ref{fig_box_plots}, which shows that the median values of the dice score distributions also steadily increase across the various configurations. Due to the low number of testing cases (n = 15), a non-parametric Wilcoxon signed-rank statistical test did not yield statistically different results. However, given the clear improvements in the Dice score, we believe that the addition of more data would provide clearer insights into any performance differences.

%% file: chapters/4_discussion_and_conclusion.tex
% ================================================
\section{Discussion and Conclusion}
\label{sec_DiscussionConclusion}
% ================================================

\noindent
In this work, we trained various configurations of the 3D nnUNet model to segment lymph nodes in mediastinal CT volumes with anatomical priors. As evidenced by prior works \cite{Liu2014_spatialPrior,Oda2018,Iuga2021_stationMapping_NIHCTLN,Bouget2022_StOlavs,Mehrtash2023}, the utilization of 28 anatomical priors ameliorated the severe class imbalance problem and reduced false positive incidences as the model had supervision in the previously uncertain anatomical regions. The 3D cascade nnUNet obtained reasonable segmentation Dice scores for all LNs and those with SAD $\geq$ 8mm; these scores tend to be lower for structures that are smaller in size in contrast to larger organs (e.g. liver). 

Presently, it is impossible for an automated approach to obtain voxel-perfect segmentations of LNs due to technical challenges in the CT acquisition process. The timing and uptake of contrast material can fluctuate, resulting in adjacent regions (e.g., azygos vein) to be iso-intense with the LNs that straddle the mediastinum as seen in Fig. \ref{fig_example_CT_predictions}(a), which can obscure their shape and size. Additionally, the manual annotations done by trained radiologists or residents for LNs in CT may not always be complete. For example, in Fig. \ref{fig_example_CT_predictions}(b), a lymph node was not annotated in the ground truth, but the nnUNet model correctly segmented this missed lymph node. Incomplete ground truth could also reduce the segmentation Dice scores as the correctly detected LN would be incorrectly considered as a false positive instead of a true positive. 

Furthermore, the true metastatic nature of a node can only be determined through an invasive biopsy procedure for diagnosis. But, this may not be clinically feasible due to small sizes or anatomic locations. Thus, reliance on CT, PET/CT, or ultrasound imaging markers are few of the non-invasive ways to assess malignancy \cite{Bouget2022_StOlavs}. Utilizing PET/CT can provide complementary information on metastatic nodes based on their metabolic activity; higher SUV values (regardless of the nodal size) are suspicious for metastatic disease. However, PET/CT is not the initial diagnostic test and is generally performed after CTs first identify a malignancy and areas of metastatic disease; to that end, the initial CT exam must be exhaustively used to derive biomarkers.  

One of the main limitations of our work is the inability to disambiguate collocated LNs in the CT volumes due to the diversity of LN shapes and appearances. As pointed out by Bouget et al. \cite{Bouget2022_StOlavs}, this task is often difficult even for an experienced radiologist, and it is expected that the task would be equally, if not more, challenging for an automated method as well. Additionally, we do not tackle the problem of station mapping in this work. Furthermore, the test dataset that we used in this work is relatively small with only 15 patients. Due to clear imbalances in the station-level distributions \cite{Bouget2022_StOlavs}, an extensive data collection and annotation process would be required to address both these issues. This would also enable any statistical differences to be extracted across the different nnUNet model configurations.

As localization of lymph nodes and measurement of suspicious nodes are routine tasks that clinicians perform on a day-to-day basis, our end-to-end anatomical prior-guided approach to segmenting lymph nodes would potentially alleviate the cumbersome nature of the measurement task. Since the models were trained with data that was presumably acquired with a variety of imaging scanners and exam protocols, it is fair to note that our 3D cascade model was particularly effective at identifying LNs with SAD $\geq$ 8mm. It holds promise as a tool to report automated measurements, differentiate metastatic from non-metastatic nodes, and flag any concerning LNs that were missed by the reading radiologist. 

In summary, the segmentation of mediastinal lymph nodes in CT was explored in our work through the use of anatomical priors. In addition to the LN labels for 89 volumes from the public NIH CT Lymph Node dataset, 28 different structures were also used to train different configuration of 3D nnUNet segmentation models in an end-to-end manner. As post-processing steps were unnecessary, our 3D cascade model was able to achieve a segmentation dice score of 72.2 $\pm$ 22.3 for clinically significant LNs with SAD $\geq$ 8mm. Our results show an improvement of 10 points over the current state-of-the-art method that was evaluated on the same test dataset. Mining additional LNs in unannotated CT exams would enable the segmentation performance to be improved over time. Our approach has immense potential for improved patient outcomes through the identification of enlarged nodes in initial staging CT exams, while also determining the best options for next steps, whether that be diagnostic biopsy or therapeutic treatment. 

%% file: chapters/5_acknowledgements.tex
% ================================================
\section*{Acknowledgments}
\label{sec_Acknowledgments}
% ================================================

\noindent
\textbf{Funding}: This work was supported by the Intramural Research Program of the NIH Clinical Center (project number 1Z01 CL040004). We also thank Jaclyn Burge for the helpful comments and suggestions.

\noindent
\textbf{Ethical approval}: All procedures performed in studies involving human participants were in accordance with the ethical standards of the institutional and/or national research committee and the 1964 Helsinki declaration and its later amendments or comparable ethical standards. For this study, informed consent was not required.

\noindent
\textbf{Conflict of Interest}: RMS receives royalties from iCAD, Philips, PingAn, ScanMed, and Translation Holdings. His lab received research support from PingAn. The authors have no additional conflicts of interest to declare. 

%% file: chapters/supplementaryMaterial.tex
% \documentclass{article}

% \usepackage[demo]{graphicx}

% \usepackage{subcaption}
% \usepackage{cleveref}

% \begin{document}
%     %% BIC and RSS distributions

% ================================================
\section{Supplementary Material}
\label{sec_supplementaryMaterial}
% ================================================

% ^^^^^^^^
% ^^^^^^^^
\begin{table*}[!ht]
\centering\fontsize{9}{12}\selectfont % to make font size 9 pt
\setlength\aboverulesep{0pt}\setlength\belowrulesep{0pt} % intersect vert and horiz lines
\setlength{\tabcolsep}{7pt} % set small spacing between entries of column (default 6pt)
\setcellgapes{3pt}\makegapedcells % small space between row entries (default 3pt)
\caption{Complete list of all organs and structures used for training the 3D nnUNet models in this work. 28 classes were generated by TotalSegmentator when it was executed on the 89 CT volumes in the NIH CT Lymph Node dataset. These were combined with the lymph node labels from the NIH CT Lymph Node dataset to yield the final 29 classes for training.}
\begin{adjustbox}{max width=\textwidth}
\begin{tabular}{@{} c|c|c|c @{}} % @ property can be modified
\toprule
\#      &   Structure                           & Class ID              & Extracted From \\
\midrule

1       &   Body Region Mask                                & 1                     & TotalSegmentator \\
2       &   Lymph Nodes                                     & 1                     & NIH CT LN data (annotated by St Olavs) \\
3       &   Spleen                                          & 3                     & TotalSegmentator \\
4       &   Kidneys (left \& right)                         & 4                     & TotalSegmentator \\
5       &   Gall Bladder                                    & 5                     & TotalSegmentator \\
6       &   Liver                                           & 6                     & TotalSegmentator \\
7       &   Stomach                                         & 7                     & TotalSegmentator \\
8       &   Aorta                                           & 8                     & TotalSegmentator \\
9       &   Inferior Vena Cava                              & 9                     & TotalSegmentator \\
10      &   Portal and Splenic Vein                         & 10                    & TotalSegmentator \\
11      &   Pancreas                                        & 11                    & TotalSegmentator \\
12      &   Adrenal Glands (left \& right)                  & 12                    & TotalSegmentator \\
13      &   Lung (all lobes)                                & 13                    & TotalSegmentator \\
14      &   \shortstack{Skeleton (vertebrae \& ribs \& pelvis \\ \& sacrum \& humerus \& scapula \\ \& clavicula \& femur \& hip)}                                                                & 14                    & TotalSegmentator \\
15      &   Esophagus                                       & 15                    & TotalSegmentator \\
16      &   Trachea                                         & 16                    & TotalSegmentator \\
17      &   Heart (Mycardium \& Atria \& Ventricles)        & 17                    & TotalSegmentator \\
18      &   Pulmonary Artery                                & 18                    & TotalSegmentator \\
19      &   Iliac Artery                                    & 19                    & TotalSegmentator \\
20      &   Iliac Vein                                      & 20                    & TotalSegmentator \\
21      &   Small Bowel                                     & 21                    & TotalSegmentator \\
22      &   Duodenum                                        & 22                    & TotalSegmentator \\
23      &   Colon                                           & 23                    & TotalSegmentator \\
24      &   Glutes Max                                      & 24                    & TotalSegmentator \\
25      &   Glutes Min                                      & 25                    & TotalSegmentator \\
26      &   Glutes Medius                                   & 26                    & TotalSegmentator \\
27      &   Autochthon                                      & 27                    & TotalSegmentator \\
28      &   Iliopsoas                                       & 28                    & TotalSegmentator \\
29      &   Urinary Bladder                                 & 29                    & TotalSegmentator \\

\bottomrule
\end{tabular}
\end{adjustbox}
\label{fig:model_selection_criteria}
\end{table*}
% ^^^^^^^^
% ^^^^^^^^

% \end{document}

%% file: main.bbl
\begin{thebibliography}{10}

\bibitem{Torabi2004}
Maha Torabi, Suzanne~L. Aquino, and Mukesh~G. Harisinghani.
\newblock Current concepts in lymph node imaging.
\newblock {\em Journal of Nuclear Medicine}, 45(9):1509--1518, 2004.

\bibitem{Ganeshalingam2009}
Skandadas Ganeshalingam and Dow-Mu Koh.
\newblock Nodal staging.
\newblock {\em Cancer Imaging}, 9(1):104--11, 2009.

\bibitem{Taupitz2007}
Matthais Taupitz.
\newblock Imaging of lymph nodes -- mri and ct.
\newblock {\em Springer}, pages 321--329, 2007.

\bibitem{Amin2017}
Mahul~B. Amin, Frederick~L. Greene, Stephen~B. Edge, Carolyn~C. Compton, Jeffrey~E. Gershenwald, Robert~K. Brookland, Laura Meyer, Donna~M. Gress, David~R. Byrd, and David~P. Winchester.
\newblock The eighth edition ajcc cancer staging manual: Continuing to build a bridge from a population-based to a more “personalized” approach to cancer staging.
\newblock {\em CA: A Cancer Journal for Clinicians}, 67(2):93--99, 2017.

\bibitem{Mao2014}
Yun Mao, Sandeep Hedgire, and Mukesh~G. Harisinghani.
\newblock Radiologic assessment of lymph nodes in oncologic patients.
\newblock {\em Curr Radiol Rep}, 2(36), 2014.

\bibitem{Liu2014_spatialPrior}
Jiamin Liu, Jocelyn Zhao, Joanne Hoffman, Jianhua Yao, Weidong Zhang, Evrim~B. Turkbey, Shijun Wang, Christine Kim, and Ronald~M. Summers.
\newblock {Mediastinal lymph node detection on thoracic CT scans using spatial prior from multi-atlas label fusion}.
\newblock In Stephen Aylward and Lubomir~M. Hadjiiski, editors, {\em Medical Imaging 2014: Computer-Aided Diagnosis}, volume 9035, page 90350M. International Society for Optics and Photonics, SPIE, 2014.

\bibitem{Roth2014_NIHCTLN}
Holger Roth, Le~Lu, Ari Seff, Kevin~M. Cherry, Joanne Hoffman, Shijun Wang, Jiamin Liu, Evrim Turkbey, and Ronald~M. Summers.
\newblock A new 2.5d representation for lymph node detection using random sets of deep convolutional neural network observations.
\newblock In {\em Medical Image Computing and Computer-Assisted Intervention – MICCAI 2014}, volume 8673, pages 520--527, 2014.

\bibitem{Oda2018}
Hirohisa Oda, Holger~R. Roth, Kanwal~K. Bhatia, Masahiro Oda, Takayuki Kitasaka, Shingo Iwano, Hirotoshi Homma, Hirotsugu Takabatake, Masaki Mori, Hiroshi Natori, Julia~A. Schnabel, and Kensaku Mori.
\newblock {Dense volumetric detection and segmentation of mediastinal lymph nodes in chest CT images}.
\newblock In Nicholas Petrick and Kensaku Mori, editors, {\em Medical Imaging 2018: Computer-Aided Diagnosis}, volume 10575, page 1057502. International Society for Optics and Photonics, SPIE, 2018.

\bibitem{Bouget2019_StOlavs_15Patients}
David Bouget, Arve Jørgensen, Gabriel Kiss, Haakon~Olav Leira, and Thomas Langø.
\newblock Semantic segmentation and detection of mediastinal lymph nodes and anatomical structures in ct data for lung cancer staging.
\newblock {\em International journal of computer assisted radiology and surgery}, 14(6):977--986, 2019.

\bibitem{Iuga2021_fovealNetwork_NIHCTLN}
Andra-Iza Iuga, Heike Carolus, Anna~J. Höink, Tom Brosch, Tobias Klinder, David Maintz, Thorsten Persigehl, Bettina Baeßler, and Michael Püsken.
\newblock Automated detection and segmentation of thoracic lymph nodes from ct using 3d foveal fully convolutional neural networks.
\newblock {\em BMC Med Imaging}, 21:69, 2021.

\bibitem{Iuga2021_stationMapping_NIHCTLN}
Andra-Iza Iuga, Tanja Lossau, Liliana~Laurenco Caldeira, Miriam Rinneburger, Simon Lennartz, Nils {Große Hokamp}, Michael Püsken, Heike Carolus, David Maintz, Tobias Klinder, and Thorsten Persigehl.
\newblock Automated mapping and n-staging of thoracic lymph nodes in contrast-enhanced ct scans of the chest using a fully convolutional neural network.
\newblock {\em European Journal of Radiology}, 139:109718, 2021.

\bibitem{Bouget2022_StOlavs}
David Bouget, André Pedersen, Johanna Vanel, Haakon~O. Leira, and Thomas Langø.
\newblock Mediastinal lymph nodes segmentation using 3d convolutional neural network ensembles and anatomical priors guiding.
\newblock {\em Computer Methods in Biomechanics and Biomedical Engineering: Imaging \& Visualization}, 11(1):44--58, 2023.

\bibitem{Mehrtash2023}
Alireza Mehrtash, Erik Ziegler, Tagwa Idris, Bhanusupriya Somarouthu, Trinity Urban, Ann~S. LaCasce, Heather Jacene, Annick~D. {Van Den Abbeele}, Steve Pieper, Gordon Harris, Ron Kikinis, and Tina Kapur.
\newblock Evaluation of mediastinal lymph node segmentation of heterogeneous ct data with full and weak supervision.
\newblock {\em Computerized Medical Imaging and Graphics}, 111:102312, 2024.

\bibitem{Lu2018_mri}
Yun Lu, Qiyue Yu, Yuanxiang Gao, Yunpeng Zhou, Guangwei Liu, Qian Dong, Jinlong Ma, Lei Ding, Hong wei Yao, Zhongtao Zhang, Gang Xiao, Qi~An, Guiying Wang, Jinchuan Xi, Wei-Tang Yuan, Yugui Lian, Dianliang Zhang, Chun-Gang Zhao, Qin Yao, Wei Liu, Xiaoming Zhou, Shuhao Liu, Qingyao Wu, Wenjian Xu, Jianli Zhang, Dong sheng Wang, Zhen qing Sun, Yuanxiang Gao, Xian xiang Zhang, Ji~lin Hu, Maoshen Zhang, Guanrong Wang, Xuefeng Zheng, Lei Wang, Jie Zhao, and Shujian Yang.
\newblock Identification of metastatic lymph nodes in mr imaging with faster region-based convolutional neural networks.
\newblock {\em Cancer research}, 78 17:5135--5143, 2018.

\bibitem{Debats2019_mri}
Oscar~A. Debats, Geert~J.S. Litjens, and Henkjan~J. Huisman.
\newblock Lymph node detection in mr lymphography: false positive reduction using multi-view convolutional neural networks.
\newblock {\em PeerJ}, 7:e8052, November 2019.

\bibitem{Mathai2021_mlmi}
Tejas~Sudharshan Mathai, Sungwon Lee, Daniel~C. Elton, Thomas~C. Shen, Yifan Peng, Zhiyong Lu, and Ronald~M. Summers.
\newblock Detection of lymph nodes in t2 mri using neural network ensembles.
\newblock In Chunfeng Lian, Xiaohuan Cao, Islem Rekik, Xuanang Xu, and Pingkun Yan, editors, {\em Machine Learning in Medical Imaging}, pages 682--691, Cham, 2021. Springer International Publishing.

\bibitem{Mathai2021_spie}
Tejas~Sudharshan Mathai, Sungwon Lee, Daniel~C. Elton, Thomas~C. Shen, Yifan Peng, Zhiyong Lu, and Ronald~M. Summers.
\newblock {Lymph node detection in T2 MRI with transformers}.
\newblock In Karen Drukker and Khan~M. Iftekharuddin, editors, {\em Medical Imaging 2022: Computer-Aided Diagnosis}, volume 12033, page 120333B. International Society for Optics and Photonics, SPIE, 2022.

\bibitem{Wang2022_mri}
Shuai Wang, Yingying Zhu, Sungwon Lee, Daniel~C. Elton, Thomas~C. Shen, Youbao Tang, Yifan Peng, Zhiyong Lu, and Ronald~M. Summers.
\newblock Global-local attention network with multi-task uncertainty loss for abnormal lymph node detection in mr images.
\newblock {\em Medical Image Analysis}, page 102345, 2022.

\bibitem{Mathai2022_CARS}
Tejas~Sudharshan Mathai, Sungwon Lee, Thomas~C. Shen, Zhiyong Lu, and Ronald~M. Summers.
\newblock Universal lymph node detection in t2 mri using neural networks.
\newblock {\em Int J CARS}, 2022.

\bibitem{Mathai2023_LNMRISeg}
Tejas~Sudharshan Mathai, Sungwon Lee, Thomas~C Shen, Daniel Elton, Zhiyong Lu, and Ronald~M Summers.
\newblock Universal detection and segmentation of lymph nodes in multi-parametric mri.
\newblock {\em International journal of computer assisted radiology and surgery}, June 2023.

\bibitem{Clark2013_TCIA}
Kenneth~W. Clark, Bruce~A. Vendt, Kirk~E. Smith, John~B. Freymann, Justin~S. Kirby, Paul Koppel, Stephen~M. Moore, Stanley~R. Phillips, David~R. Maffitt, Michael Pringle, Lawrence Tarbox, and Fred~W. Prior.
\newblock The cancer imaging archive (tcia): Maintaining and operating a public information repository.
\newblock {\em J. Digital Imaging}, 26(6):1045--1057, 2013.

\bibitem{Reynisson2015_StOlavs_lungCancerPatientsOriginalData}
Pall~Jens Reynisson, Marta Scali, Erik Smistad, Erlend~Fagertun Hofstad, Håkon~Olav Leira, Frank Lindseth, Toril~Anita Nagelhus~Hernes, Tore Amundsen, Hanne Sorger, and Thomas Langø.
\newblock Airway segmentation and centerline extraction from thoracic ct – comparison of a new method to state of the art commercialized methods.
\newblock {\em PLOS ONE}, 10(12):1--20, 12 2015.

\bibitem{Wasserthal2023_TS}
Jakob Wasserthal, Hanns-Christian Breit, Manfred~T. Meyer, Maurice Pradella, Daniel Hinck, Alexander~W. Sauter, Tobias Heye, Daniel~T. Boll, Joshy Cyriac, Shan Yang, Michael Bach, and Martin Segeroth.
\newblock Totalsegmentator: Robust segmentation of 104 anatomic structures in ct images.
\newblock {\em Radiology: Artificial Intelligence}, 5(5):e230024, 2023.

\bibitem{Hou2024_bodyComposition}
Benjamin Hou, Tejas~Sudharshan Mathai, Jianfei Liu, Christopher Parnell, and Ronald~M. Summers.
\newblock Enhanced muscle and fat segmentation for ct-based body composition analysis: A comparative study.
\newblock {\em arXiv}, 2024.

\bibitem{Lee2023_bodyCompositionAnalysis}
Matthew~H. Lee, Daniel Liu, John~W. Garrett, Alberto Perez, Ryan Zea, Ronald~M. Summers, and Perry~J. Pickhardt.
\newblock Comparing fully automated ai body composition measures derived from thin and thick slice ct image data.
\newblock {\em Abdominal Radiology}, pages 1--1, 2023.

\bibitem{Isensee_2021}
Fabian Isensee, Paul~F Jaeger, Simon A~A Kohl, Jens Petersen, and Klaus~H Maier-Hein.
\newblock {nnU-Net}: a self-configuring method for deep learning-based biomedical image segmentation.
\newblock {\em Nat. Methods}, 18(2):203--211, February 2021.

\end{thebibliography}
